\documentclass[conference]{worldcomp}

\usepackage[hmargin=.75in,vmargin=1in,left=0.7in,right=0.7in]{geometry}
\usepackage[american]{babel}
\usepackage[T1]{fontenc}
\usepackage{times}
\usepackage{caption}

\usepackage{textcomp}
\usepackage{epsfig,graphicx}
\usepackage{xcolor}
\usepackage{amsfonts,amsmath,amssymb}
\usepackage{fixltx2e} % Fixing numbering problem when using figure/table* 
\usepackage{booktabs}

\usepackage{stmaryrd}
\usepackage{float}
\usepackage{algorithm}
\usepackage{algcompatible}
\usepackage{xspace}

\newtheorem{theorem}{Theorem}

\newtheorem{example}[theorem]{Example}
\newtheorem{lemma}[theorem]{Lemma}

\algnewcommand\AND{\textbf{and}\xspace}
\algnewcommand\OR{\textbf{or}\xspace}

\title{\bf Node-Independent Spanning Trees in Gaussian Networks}           %%%% Replace with your title.

\author{
{\bfseries Z. Hussain$^1$, B. AlBdaiwi$^1$, and A. Cerny$^2$}\\
$^1$Computer Science Department, Kuwait University, Kuwait\\
$^2$Department of Information Science, Kuwait University, Kuwait\\
}

\begin{document}

\maketitle                        %%%% To set Title and Author names.

\begin{abstract}
Message broadcasting in networks could be carried over spanning trees. A set of spanning trees in the same network is node independent if two conditions are satisfied. First, all trees are rooted at node $r$. Second, for every node $u$ in the network, all trees' paths from $r$ to $u$ are node-disjoint, excluding the end nodes $r$ and $u$. Independent spanning trees have applications in fault-tolerant communications and secure message distributions.

Gaussian networks and two-dimensional toroidal networks share similar topological characteristics. They are regular of degree four, symmetric, and node-transitive. Gaussian networks, however, have relatively lesser network diameter that could result in a better performance. This promotes Gaussian networks to be a potential alternative for two-dimensional toroidal networks.

In this paper, we present constructions for node independent spanning trees in dense Gaussian networks. Based on these constructions, we design routing algorithms that can be used in fault-tolerant routing and secure message distribution.
We also design fault-tolerant algorithms to construct these trees in parallel.
\end{abstract}

\vspace{1em}

\section{Introduction}\label{SectionIntroduction}
A parallel computing system interconnection topology decides the system fault-tolerance capabilities as well as the system over all communication efficiency. There are many researches on interconnection networks such as hypercube \cite{hayes1989hypercube}\cite{hayes1986architecture}, generalized hypercube \cite{bhuyan1984generalized}, mesh, $k$--ary $n$--cube \cite{bose1995lee}, torus \cite{dally1986torus}, REFINE \cite{Bhandarkar19951783}, and RMRN \cite{RMRN}\cite{Bhandarkar1995107} networks; more examples of interconnection netowkrs can be found in \cite{Duato:1997:INE:550183}\cite{Kumar:2003:IPC:600009}.

An efficient interconnection topology called Gaussian network has been studied in~\cite{4483506}\cite{Martinez06densegaussian}.
Gaussian networks and two-dimensional toroidal networks share similar topological properties. They are symmetric, node-transitive, and regular degree four. Gaussian networks, however, have relatively lesser diameter which suggests that they could
be a potential alternative for two-dimensional toroidal networks. Gaussian networks have been studied in \cite{bose2015higher}\cite{5204080}\cite{hussain2015better}, and they are briefly reviewed in Section \ref{SectionBackground}.

In parallel computing as well as in distributed systems, a network can be represented as a graph where nodes and edges represent processors and communication links among the processors, respectively. A path from a node $s$ to a node $d$ in a graph is a sequence of edges, which connects a sequence of nodes from $s$ to $d$. Two such paths are said to be independent if their nodes are disjoint except for the end nodes $s$ and $d$, i.e. there are no common intermediate nodes in the two paths. A spanning tree is a connected loop-free subgraph of graph $G$ containing all the nodes of graph $G$. Spanning trees rooted at node $r$ are said to be independent if the paths from $r$ to any other node $u$ in any two of the trees are independent. 

Node independent spanning trees used to resolve important issues in network applications such as fault-tolerant broadcasting \cite{Itai:1988:MAR:52722.52725}\cite{Krishnamoorthy:1987:FDI:35064.36256} and secure message distribution \cite{Rescigno2001}\cite{Yang20111254}. These applications are briefly described below:
\begin{itemize}
\item Consider the existence of $t$ node independent spanning trees rooted at node $r$ in network $N$. Assume that $N$ contains at most $t-1$ faulty nodes. Then, $r$ can broadcast a message to every non-faulty node $u$ in $N$ by broadcasting the message over all the~$t$ trees.
Since the number of faulty nodes is less than $t$, at least one of the $t$ node disjoint paths from $r$ to $u$ is fault free.
\item Node independent spanning trees could be used in secure message distribution over
a fault-free network as follows.
A message can be divided into $t$ packets where each packet is sent by node $r$ to its destination using a different spanning tree. Thus, each node in the network receives at most one of the $t$ packets except
the destination node that receives all the $t$ packets \cite{Lin2010414}\cite{Yang20111254}\cite{Yang:2009:IST:1726593.1728973}.
\end{itemize}

In \cite{AlBdaiwi2016}, AlBdaiwi et. al. two edge-disjoint node-independent spanning trees were constructed in dense Gaussian network. Such network contains the maximum number of nodes for a given diameter $k$ \cite{Martinez06densegaussian}, where the depth of each tree is $2k$, $k \geq 1$.
Based on those constructions, algorithms were provided for fault-tolerant routing and secure message distribution where the source node can be any node in the network.

In \cite{4ISTConference}, we gave algorithms for constructing four node independent spanning trees in a dense Gaussian network of depth $k$ working in $2k$ steps where the trees are not necessarily edge-disjoint. In this paper, we extend our work and present a $k$--steps construction algorithm. Furthermore, we develop routing algorithms based on these trees that
can be used in fault-tolerant communication and secure message distribution.

The rest of this paper is organized as follows. The Gaussian network is reviewed in Section \ref{SectionBackground}. In Section \ref{SectionTheFourTrees}, we introduce a construction of
 four node independent spanning trees in Gaussian networks.
Based on the constructed trees, routing algorithms from a given source node to a given destination node are described in Section \ref{SectionRouting}.
Section \ref{SectionKConstruction} presents parallel algorithms for constructing the four
node independent spanning trees, and Section \ref{SectionSimulation} presents simulation
analysis for these algorithms.
Finally, The paper is concluded in Section \ref{SectionConcl}.

\section{Background\label{SectionBackground}}

Gaussian networks are regular symmetric networks of degree 4.
These networks are based on quotient rings of Gaussian integers $\mathbb{Z}[i] = \{x+yi \mid x,y \in \mathbb{Z}\}$ where $i = \sqrt{-1}$ \cite{272484}. $\mathbb{Z}[i]$ is an Euclidean domain. The nodes of the network are elements of the residue class modulo some $\alpha \in \mathbb{Z}[i]$, where $\alpha$ is considered as a network generator. The norm $N(\alpha) = a^2 + b^2$ of a Gaussian network generated by $\alpha = a + bi$, is the total number of nodes in this network.
A dense diameter-optimal Gaussian network denoted by $G(\alpha_k)$, where $\alpha_k = k+(k + 1)i$,
is a $k$ diameter Gaussian network that contains the maximum number of nodes. Note that
$\alpha_k$ always generates the maximum possible number of nodes within
diameter $k$ since GCD$(k, k+1) = 1$.

\begin{figure}[ht]
    \centering
    \includegraphics[scale=0.9]{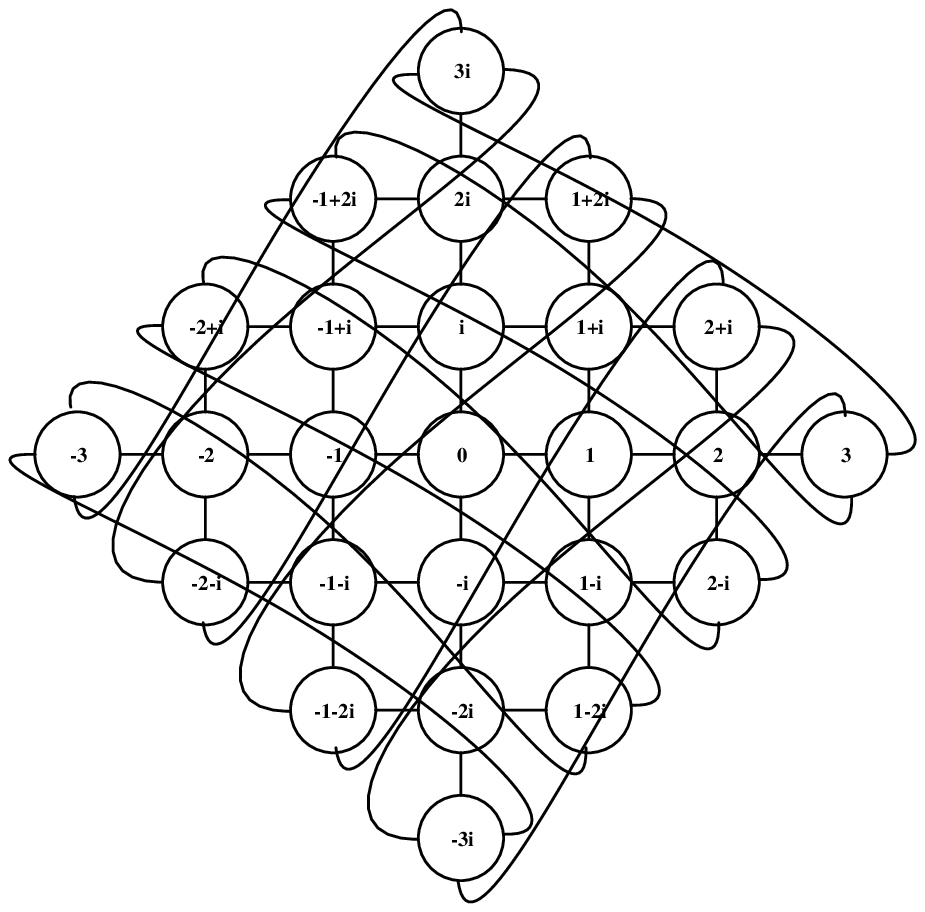}
    \caption{Gaussian network generated by $\alpha = 3+4i$}
    \label{3+4i}
\end{figure}

Gaussian networks can be represented in several different ways as described in \cite{10.2307/2688007}.
In this paper, we deal with  graphs isomorphic to dense diameter-optimal Gaussian networks. We denote $G_k = (V_k,E_k) = G(\alpha_k)$, $k \geq 1$.
Each node in the graph is labeled as $x+yi$ where $|x| + |y| \leq k$. Two nodes $A$ and $B$ are adjacent if and only if $(A - B) \ mod \ \alpha$ is equal to $\pm 1$ or $\pm i$. As described in \cite{4483506}, the number of nodes at distance $j$ from any single node in a dense Gaussian network is $D(j)= 4j$ for $j = 1, 2, \dots, k$. Figure \ref{3+4i} shows an example of Gaussian network generated by $\alpha = 3+4i$.

\begin{figure}[ht]
    \centering
    \includegraphics[scale=0.65]{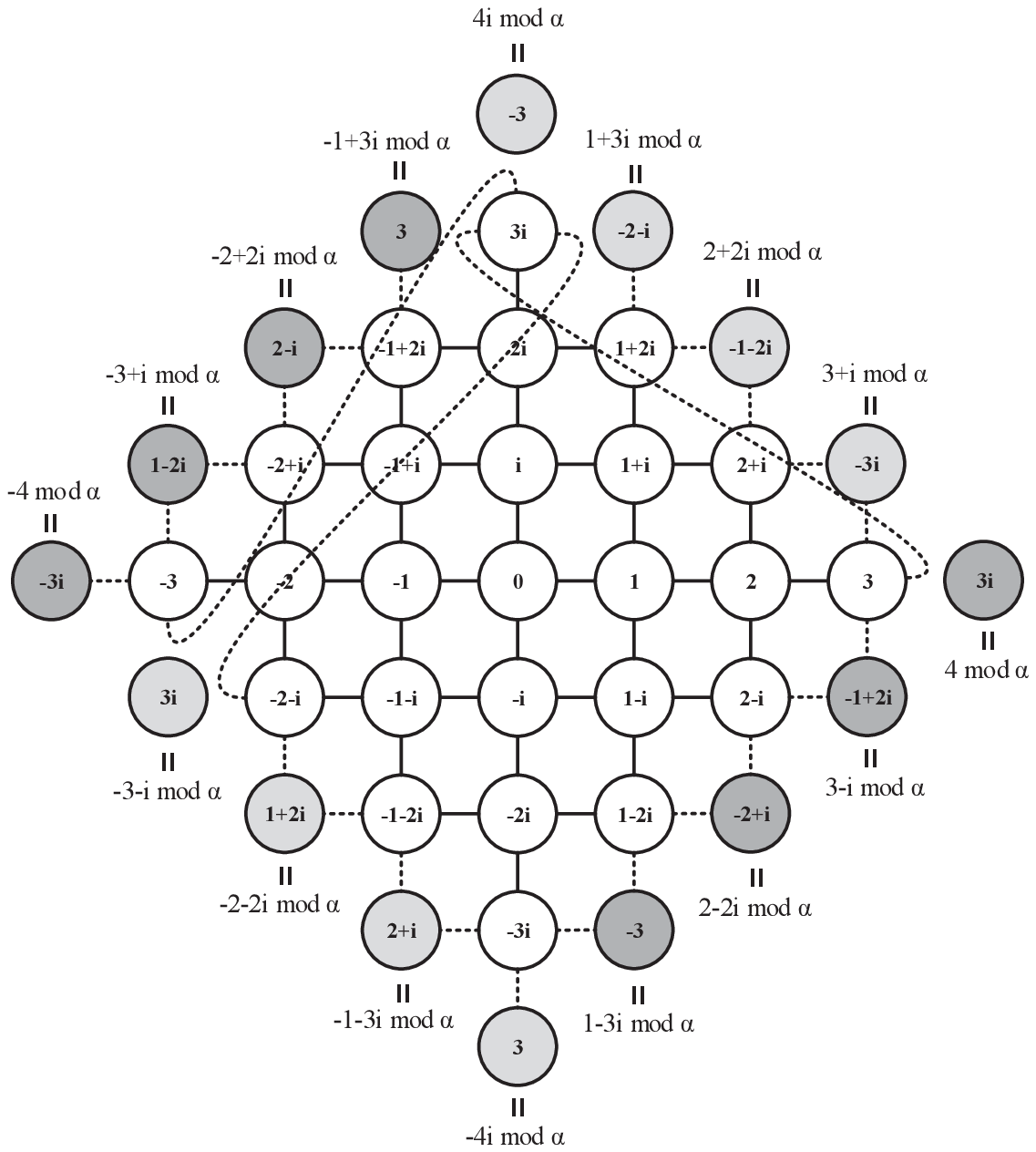}
    \caption{Wraparound links in Gaussian network generated by $\alpha = 3+4i$}
    \label{3+4i_wraparound}
\end{figure}

The wraparound edges in Figure \ref{3+4i} can be illustrated as straight lines by tiling the Gaussian network on an infinite grid as depicted in Figure \ref{3+4i_wraparound}. Consider the boundary node $3i$. Its $-1$, $+i$, and $+1$ edges are wraparound links that are connected to the boundary nodes $3$, $-3$, and $-2-i$, respectively. We have kept these edges as wraparounds to show the equivalence of the two illustrations. Note that the different gray color nodes are related to different network tiles. 

\section{Spanning Trees in $G_{k}=(V_{k},E_{k})$\label{SectionTheFourTrees}}
%The preliminary results of this section were presented in~\cite{4ISTConference}.
In this section we describe four spanning trees 
rooted at node $0$ in $G_{k}=\left(  V_{k},E_{k}\right)  $, $k\geq 2$,
and we will show that they are pairwise node independent in Section \ref{SectionRouting}. 
Any two spanning trees rooted at $0$ are node independent when $k=1$.
However, a motivation of our research is to tolerate transient node failures,
and thus it is assumed that $k\geq 2$.

\begin{lemma}
In $G_{k}$, $k\geq 2$, there are at most four pairwise node independent
spanning trees rooted at node $0$.
\end{lemma}

\begin{proof}
Let there be five such trees. Consider a node $v$ at distance $2$ from the
node $0$. In each of the trees there is a path of length at least 2 from the
root $0$ to $v$. By the pigeonhole principle, there are two of the trees where
such path starts by the same edge. Such two trees are not node-independent.
\end{proof}

To define our four trees, we will introduce some auxiliary notations. Recall
that $\alpha_{k}=k+\left(  k+1\right)  \mathbf{i}$. Let $E_{k}^{V}=\left\{
\left(  v,\left(  v+\mathbf{i}\right)  \operatorname{mod}\alpha_{k}\right)
;v\in V_{k}\right\}  $ denote the set of all vertical edges in
$E_{k}$. Clearly, $\left\vert E_{k}^{V}\right\vert =\left\vert V_{k}%
\right\vert $. We define the counter clockwise 90$^{\text{o}}$ rotation mapping $\rho
$\ on $V_{k}$ as $\rho\left(  x+y\mathbf{i}\right)  =\left(  -y+x\mathbf{i}%
\right)  $. The rotation $\rho$ is a bijection on $V_{k}$. We extend this
mapping to the set $E_{k}$; for $\left(  u,v\right)  \in E_{k}$ we define
$\rho\left(  u,v\right)  =\left(  \rho\left(  u\right)  ,\rho\left(  v\right)
\right)  $. For a subgraph $G^{\prime}=\left(  V^{\prime},E^{\prime}\right)  $
of $G_{k}$ we denote $\rho\left(  G^{\prime}\right)  =\left(  \rho\left(
V^{\prime}\right)  ,\rho\left(  E^{\prime}\right)  \right)  $.

\begin{figure}[!ht]
\centering
\includegraphics[scale=0.7]{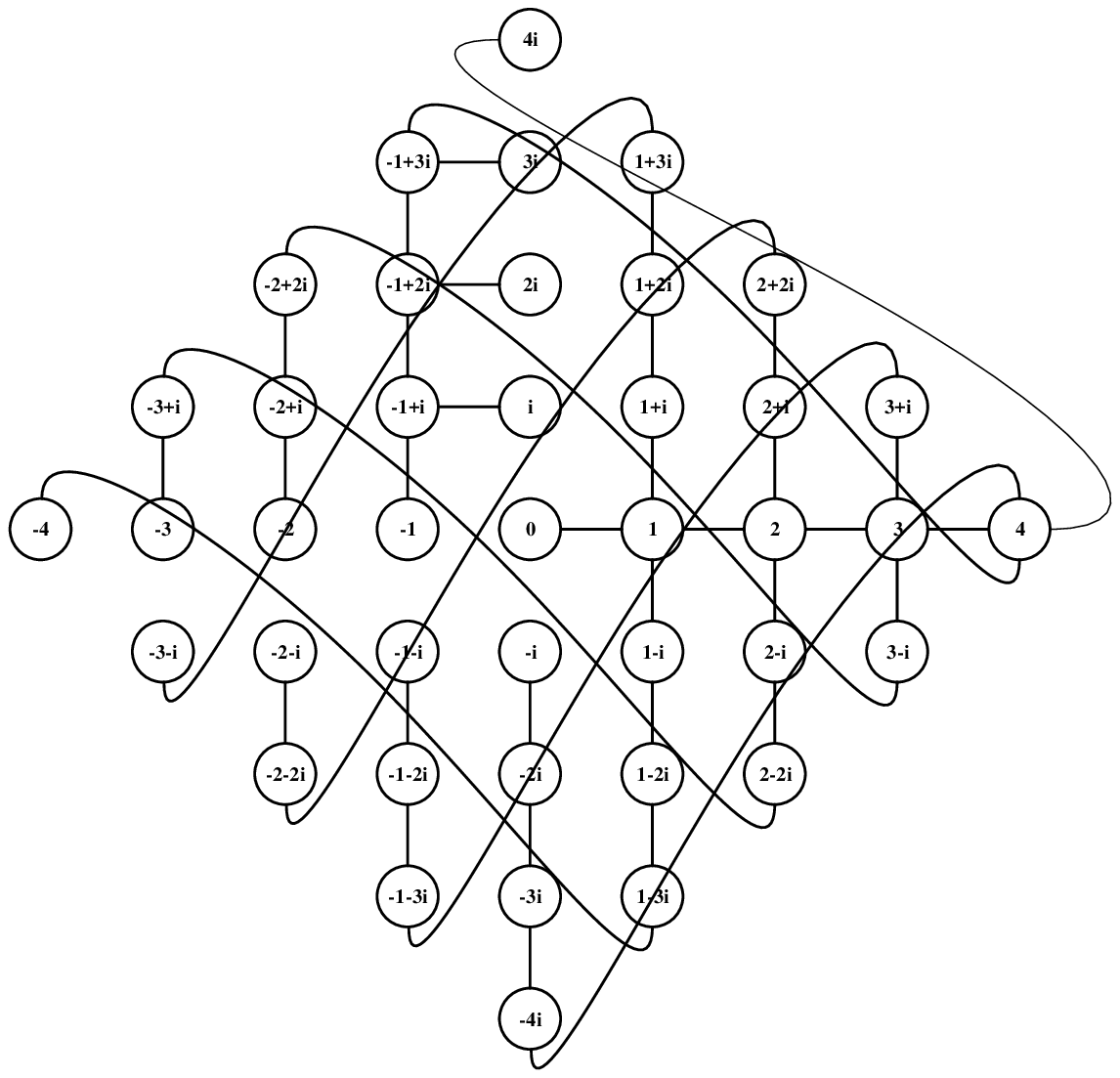} \caption{The tree $T_1,k=4$} %
\label{T1}%
\end{figure}

\begin{figure}[!ht]
\centering
\includegraphics[scale=0.7]{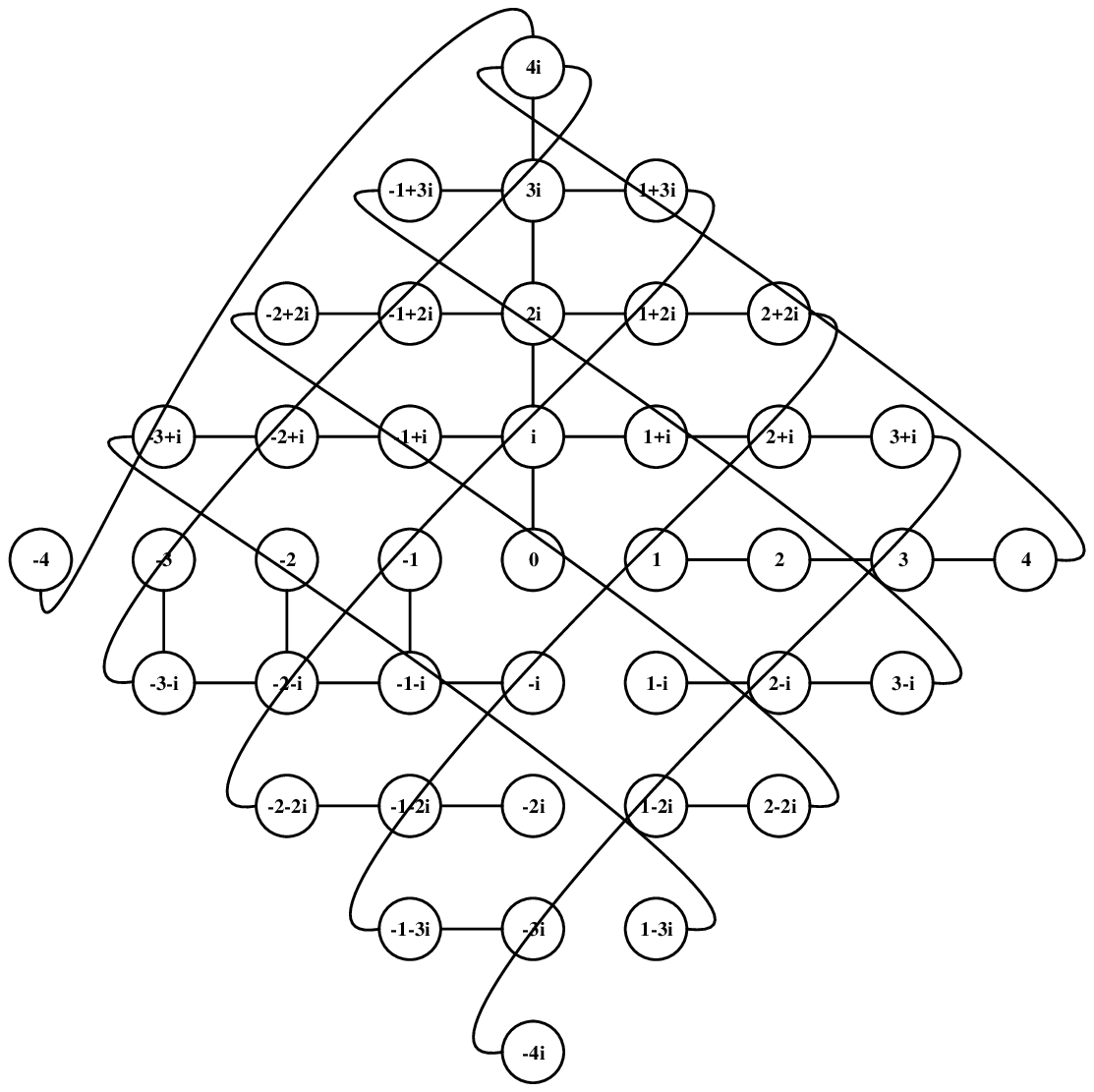} \caption{The tree $T_2,k=4$} %
\label{T2}%
\end{figure}

\begin{figure}[!ht]
\centering
\includegraphics[scale=0.7]{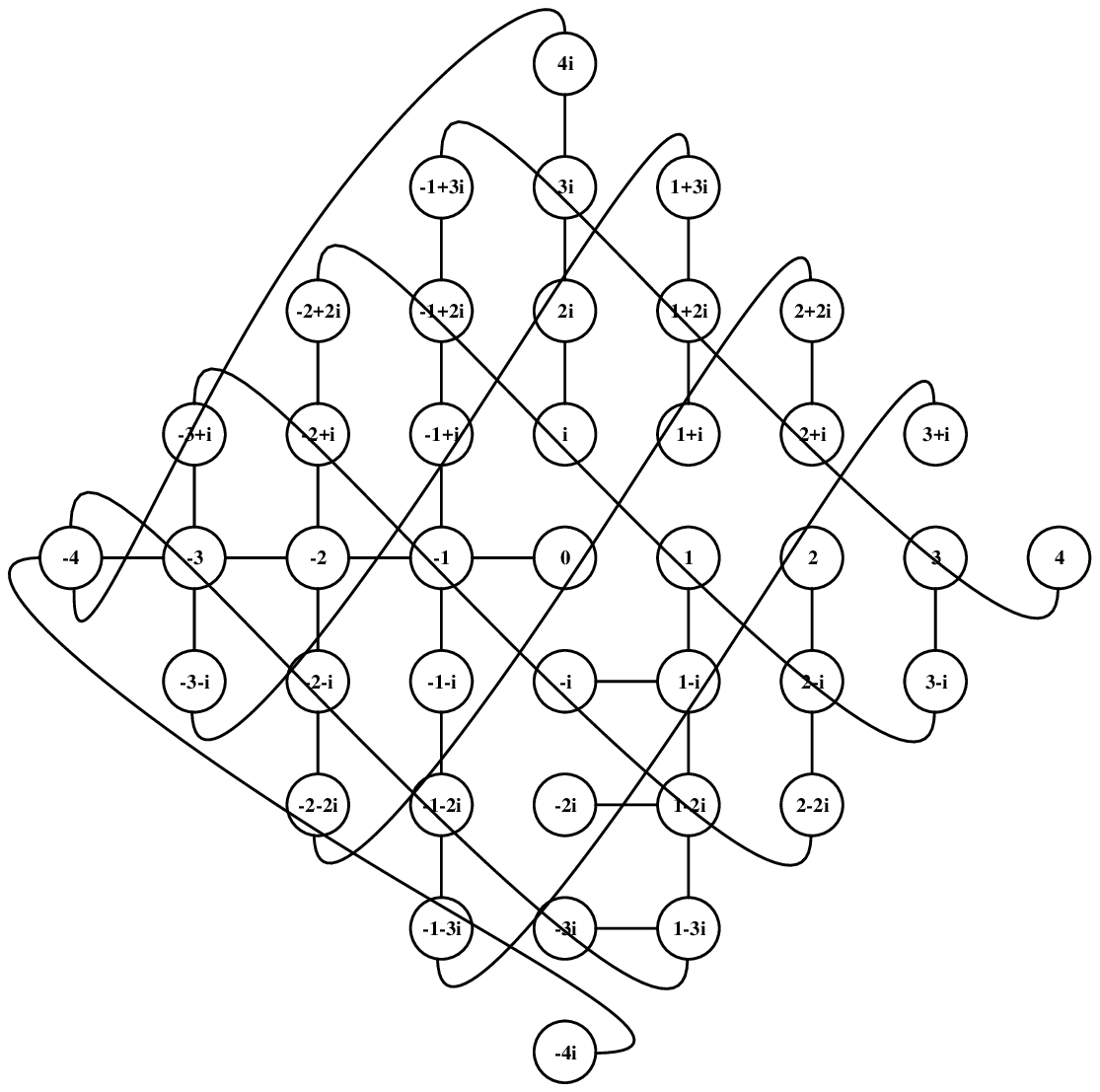} \caption{The tree $T_3,k=4$} %
\label{T3}%
\end{figure}

\begin{figure}[!ht]
\centering
\includegraphics[scale=0.7]{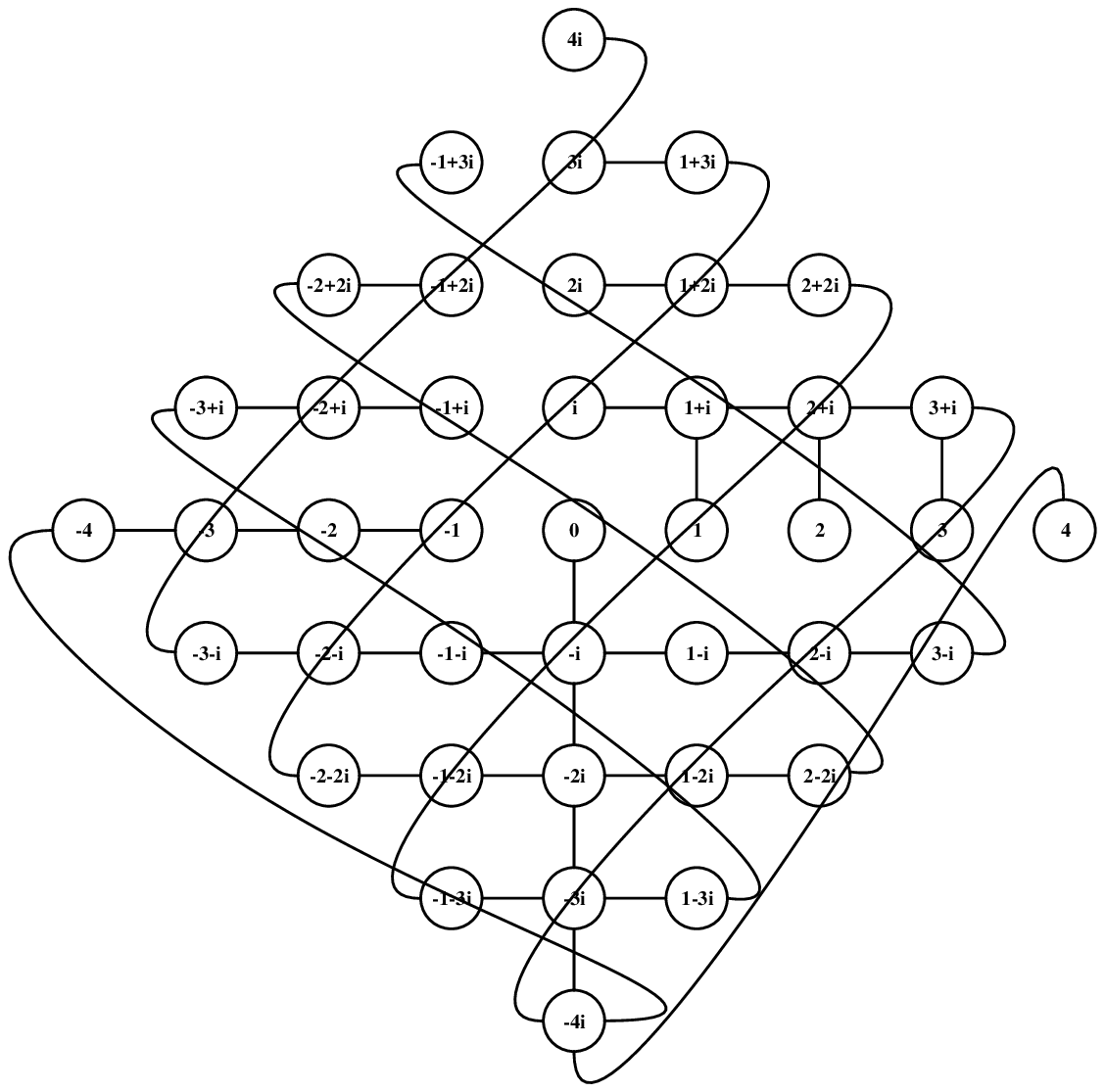} \caption{The tree $T_4,k=4$} %
\label{T4}%
\end{figure}

Now we are ready to describe four trees $\ T_{k}^{\left(  j\right)  }=\left(
V_{k}.E_{k}^{\left(  i\right)  }\right)  ,j=1,2,3,4$ - see Figures
\ref{T1}--\ref{T4}. In fact, we will define the tree $T_{k}^{\left(  1\right)  }$
only, since the remaining three trees can be described as $T_{k}^{\left(
j\right)  }=\rho^{j-1}\left(  T_{k}^{\left(  1\right)  }\right)  ,j=2,3,4$. It
is therefore sufficient to define the edge set $E_{k}^{\left(  1\right)  }$ as follows:
\[%
\begin{tabular}
[c]{rl}%
$E_{k}^{\left(  1\right)  }=$ & $E_{k}^{V}-\left\{  \left(  qi\mathbf{,}%
\left(  q+1\right)  i|0\leq q\leq k-1\right)  \right\}  $\\
& $-\left\{  \left(  q-1,q\right)  ;-k+1\leq q\leq0\right\}  $\\
& $-\left\{  \left(  -k,k\mathbf{i}\right)  \right\}  $\\
& $\cup\left\{  \left(  q,q+1\right)  |0\leq q\leq k-1\right\}  $\\
& $\cup\left\{  \left(  -1+qi,qi\right)  |1\leq q\leq k-1\right\}  $\\
& $\cup\left\{  \left(  k,ki\right)  \right\}  $%
\end{tabular}
\
\]

\begin{lemma}
$T_{k}^{\left(  j\right)  },j=1,2,3,4$ are spanning trees in $G_{k}$.
\end{lemma}

\begin{proof}
Since $\rho$ is a bijection, it is enough to prove that $T_{k}^{\left(
1\right)  }$ is a spanning tree. Based on the definition of $E^{\left(
1\right)  }$, the graph $T_{k}^{\left(  1\right)  }$ consists of $\left\vert
E_{k}^{V}\right\vert -k-k-1+k+\left(  k-1\right)  +1=\left\vert V_{k}%
\right\vert -1$ edges. We will show that $T_{k}^{\left(  1\right)  }$ is
connected by showing that there is a path in $T_{k}^{\left(  1\right)  }$ from
the node $0$ to each other node $v\in V_{k}.$ We will describe the path by a
word on the alphabet $\left\{  1,-1,i,-i\right\}  ,$ the symbols denoting the
direction of the edges to be passed. The paths are described in
Table~\ref{T1Steps}, where the node $v=c+di$, for the values satisfying $\left\vert
c\right\vert +\left\vert d\right\vert \leq k$, is considered.%
\begin{table}[H]
\centering
\caption{Steps from node 0 to node $v=c+di$ in $T_{k}^{\left(  1\right)  }$}
\label{T1Steps}
\begin{tabular}
[c]{|l|l|l|}\hline
\multicolumn{2}{|l|}{Node $c+di\neq0$} & Path\\\hline
$1\leq c\leq k-1$ & $1\leq d\leq k-c$ & $1^{c}i^{d}$\\\hline
$c=0$ & $d=k$ & $1^{k+1}$\\\hline
$c=0$ & $1<d\leq k-1$ & $1^{k}\left(  -i\right)  ^{k-d}1$\\\hline
$-k\leq c\leq-1$ & $0\leq d\leq k+c$ & $1^{k+c+1}\left(  -i\right)  ^{k-d}%
$\\\hline
$-k+1\leq c\leq0$ & $-k-c\leq d\leq-1$ & $1^{k+c}i^{k+d+1}$\\\hline
$1\leq c\leq k$ & $-k+c\leq d\leq0$ & $1^{c}\left(  -i\right)  ^{-d}$\\\hline
\end{tabular}
\end{table}
\end{proof}

\begin{lemma}
\label{height}
The height of each of the trees $T_{k}^{\left(  j\right)  }$ is $2k$, $j=1,2,3,4$.
\end{lemma}

\begin{proof}
Based on Table~\ref{T1Steps}, the longest path in $T_{k}^{\left(  1\right)  }%
$starting from the root $0$ of length $2k$ leads to node $i$ or to node
$-1$.
\end{proof}

\begin{example}
The path from the root $0$ to the node $-2+2i$ in $T_{4}^{\left(  1\right)  }$
is described as $1^{4-2+1}\left(  -i\right)  ^{4-2}=1^{3}\left(  -i\right)
^{2}$. The path is the sequence%
\begin{equation}
0,1,2,3,3-i,-2+2i\text{.} \label{T1path}%
\end{equation}

The path from the root $0$ to the node $-2-2i$ in $T_{4}^{\left(  2\right)  }$
can be obtained the following way. Since $T_{k}^{\left(  2\right)  }%
=\rho\left(  T_{4}^{\left(  1\right)  }\right)  $, we have to observe the path
in $T_{k}^{\left(  1\right)  }$ from the root $0$ to the node $\rho
^{-1}\left(  -2-2i\right)  =-2+2i,$ which is the path (\ref{T1path}). The
required path in $T_{k}^{\left(  2\right)  }$ is obtained from (\ref{T1path}) by
rotating each node using the mapping $\rho$. The resulting path is
\[
0,i,2i.3i,1+3i,-2-2i\text{.}%
\]

\end{example}

\section{Routing Using Node-Independent Spanning Trees\label{SectionRouting}}
In Section \ref{SectionTheFourTrees}, we defined four spanning trees of the
graph $G_{k}$ and in Section \ref{SectionKConstruction} we give algorithms for
their construction. In this section we present routing algorithms for
delivering messages along any of the four trees.

Routing a message consists of three types of nodes: source node, transient
nodes, and destination node. The source node is the sender who sends a message
to a destination node. A transient node is an
intermediate node that forwards the message toward the destination node. We will
provide a routing algorithm for delivering a message virtually from any node
$s$ to any node $d$ in $G_{k}$ along any of the four trees rooted at $r$. We
will provide the routing decision in any transient node $t$ on the path from
$s$ to $d$ and we will show that such decision can be taken uniquely without
information on which of the four trees is involved. This fact implies the
independence of the four trees.

Let $s\in V_{k}$. Considering the automorphism $\tau_{s}$ on $G_{k}$ defined
for $x\in V_{k}$ as $\tau_{s}\left(  x\right)  =\left(  x+s\right)
\operatorname{mod}\alpha_{k}$ and for $\left(  x,y\right)  \in$\ $E_{k}$ as
$\tau_{s}\left(  \left(  x,y\right)  \right)  =$ $\left(  \tau_{s}\left(
x\right)  ,\tau_{s}\left(  x\right)  \right)  $ allows us to describe routing
for the case $s=0$ only. The routing decisions for messages sent from a node
$s\neq0$ to some\ node $d$ is then implied by the routing decisions for
messages sent from node $0$ to node $\left(  d-s\right)
\operatorname{mod}\alpha_{k}$.

\begin{figure}[H]
    \centering
    \includegraphics[scale=0.7]{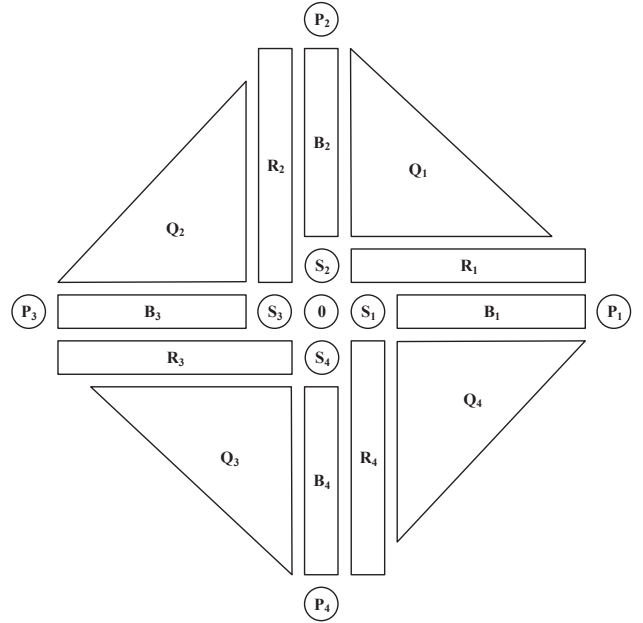}
    \caption{The partitions of Gaussian network}
    \label{GaussianPartitions}
\end{figure}

To simplify the routing process, we partition the Gaussian network into disjoint subsets
as follows (see Figure \ref{GaussianPartitions}):
%To easily construct the four independent spanning trees in $k$ steps we have categorized the nodes of the Gaussian network based on their locations. The locations are described as sets - partitions (see Figure \ref{GaussianPartitions}) defined as subsets of $V_{k}$ (i.e., the node $x+yi$ in each of the sets satisfies $\vert x \vert + \vert y \vert \leq k$). They are given as:
\newline
$B_1=\{x+yi \mid 1 < x < k, y = 0\}$, \newline
$R_1=\{x+yi \mid k > x > 0, y = 1\}$, \newline
$Q_1=\{x+yi \mid x > 0, y > 1, x + y \le k\}$, \newline
$P_1=\{x+yi \mid x = k, y = 0\}$, \newline
$S_1=\{x+yi \mid x = 1, y = 0\}$, \newline
for $j \in \{2,3,4\}$: \newline
$B_j=\rho^{-j+1}(B_1)$, $R_j=\rho^{-j+1}(R_1)$, $Q_j=\rho^{-j+1}(Q_1)$, $P_j=\rho^{-j+1}(P_1)$,
and $S_j=\rho^{-j+1}(S_1)$.

Based on Figure \ref{GaussianPartitions}, Table \ref{routingDirection} shows the
direction to which a message sent from the source node $0$ is forwarded by a
transient node $t=t_{x}+t_{y}i$, based on the location of the destination node
$d=d_{x}+d_{y}i\neq t$. The left column denotes the location of the current
transient node, the remaining columns denote the location of the destination
node. The direction is described by one of the elements in $\{1,-1,i,-i\}$.
The superscripts in brackets are to be understood as comments denoting which
tree the message is passed along. To determine the direction, in several cases
an additional condition is to be evaluated. The direction is then the result
of one of the conditional statements listed in the table
\ref{routingConditions}, denoted as $C_{1},\dots,C_{8}$.

\begin{table}[ptb]
\caption{Message routing direction}%
\label{routingDirection}
\centering
\begin{tabular}
[c]{l|c|c|c|c|}\cline{2-5}
& $S_{1}\cup B_{1}$ & $R_{1}$ & $Q_{1}$ & $P_{1}$\\\hline
\multicolumn{1}{|c|}{$S_{1}\cup B_{1}$} & $C_{1}$ & $C_{2}$ & $C_{2}$ &
$1^{\left[  1\right]  }$\\\hline
\multicolumn{1}{|c|}{$R_{1}$} & $C_{3}$ & $C_{4}$ & $i^{\left[  1\right]  }$ &
--\\\hline
\multicolumn{1}{|c|}{$Q_{1}$} & -- & $-i^{\left[  3\right]  }$ & $C_{5}$ &
--\\\hline
\multicolumn{1}{|c|}{$P_{1}$} & $-i^{\left[  2\right]  }$ & -- & -- &
--\\\hline
\multicolumn{1}{|c|}{$S_{2}\cup B_{2}$} & $i^{\left[  2\right]  }$ & $C_{6}$ &
$C_{6}$ & $i^{\left[  2\right]  }$\\\hline
\multicolumn{1}{|c|}{$R_{2}$} & -- & -- & -- & $i^{\left[  3\right]  }%
$\\\hline
\multicolumn{1}{|c|}{$Q_{2}$} & $i^{\left[  3\right]  }$ & -- & -- &
--\\\hline
\multicolumn{1}{|c|}{$P_{2}$} & $-1^{\left[  2\right]  }$ & -- & -- &
$-1^{\left[  2\right]  }$\\\hline
\multicolumn{1}{|c|}{$S_{3}\cup B_{3}$} & $C_{7}$ & $C_{7}$ & $C_{7}$ &
$i^{\left[  3\right]  }$\\\hline
\multicolumn{1}{|c|}{$R_{3}$} & -- & $-i^{\left[  3\right]  }$ & $-i^{\left[
3\right]  }$ & --\\\hline
\multicolumn{1}{|c|}{$Q_{3}$} & -- & $-i^{\left[  3\right]  }$ & $-i^{\left[
3\right]  }$ & --\\\hline
\multicolumn{1}{|c|}{$P_{3}$} & $i^{\left[  3\right]  }$ & -- & -- &
--\\\hline
\multicolumn{1}{|c|}{$S_{4}\cup B_{4}$} & $-i^{\left[  4\right]  }$ &
$-i^{\left[  4\right]  }$ & $C_{8}$ & $-i^{\left[  4\right]  }$\\\hline
\multicolumn{1}{|c|}{$R_{4}$} & -- & -- & -- & --\\\hline
\multicolumn{1}{|c|}{$Q_{4}$} & $i^{\left[  3\right]  }$ & -- & -- &
--\\\hline
\multicolumn{1}{|c|}{$P_{4}$} & $-1^{\left[  4\right]  }$ & $-1^{\left[
4\right]  }$ & -- & $-i^{\left[  4\right]  }$\\\hline
\end{tabular}
\end{table}

\begin{table}[ptb]
\caption{Conditional moves in \newline message routing direction}%
\label{routingConditions}
\centering
\begin{tabular}
[c]{|r|lll|}\hline
$C_{1}$ & If $t_{x}<d_{x}$ & then & $1^{\left[  1\right]  }$\\
&  & else & $-1^{\left[  2\right]  }$\\\hline
$C_{2}$ & If $t_{x}=d_{x}$ & then & $i^{\left[  1\right]  }$\\
&  & else & $1^{\left[  1\right]  }$\\\hline
$C_{3}$ & If $t_{x}=d_{x}$ & then & $-i^{\left[  4\right]  }$\\
&  & else & $-1^{\left[  4\right]  }$\\\hline
$C_{4}$ & If $t_{x}<d_{x}$ & then & $1^{\left[  2\right]  }$\\
&  & else & $-1^{\left[  4\right]  }$\\\hline
$C_{5}$ & If $t_{x}=d_{x}$ & then & \\
&  & If $t_{y}<d_{y}$ & then $i^{\left[  1\right]  }$\\
&  &  & else $-i^{\left[  3\right]  }$\\
&  & else & \\
&  & If $t_{x}<d_{x}$ & then $1^{\left[  2\right]  }$\\
&  &  & else $-1^{\left[  4\right]  }$\\\hline
$C_{6}$ & If $t_{y}=d_{y}$ & then & $1^{\left[  2\right]  }$\\
&  & else & $i^{\left[  2\right]  }$\\\hline
$C_{7}$ & If $t_{x}=-k-1+d_{x}$ & then & $i^{\left[  3\right]  }$\\
&  & else & $-1^{\left[  3\right]  }$\\\hline
$C_{8}$ & If $t_{y}=-k-1+d_{y}$ & then & $-1^{\left[  4\right]  }$\\
&  & else & $-i^{\left[  4\right]  }$\\\hline
\end{tabular}
\end{table}

Table \ref{routingDirection} describes the action of any transient node for
the cases when the destination node is in one of the areas from $B_{1}$,
$R_{1}$, $Q_{1}$, $P_{1}$, $S_{1}$, only. For the remaining cases, the action
is obtained by applying the proper rotation.

\begin{example}
Assume we want to determine the action of a transient node from $B_{2}$ given that
the destination node is in $Q_{4}$. Since $Q_{4}=\rho^{3}(Q_{1})$, we have to
observe the action described in our table in the row $S_{3}\cup B_{3}$ and
column $\rho^{-3}(Q_{4})=Q_{1}$. The action is determined by the conditional
statement $C_{7}$. Stated as \textquotedblleft If $t_{x}^{\prime}%
=-k-1+d_{x}^{\prime}$ then $-i^{\left[  3\right]  }$ else $-1^{\left[
3\right]  }$\textquotedblright, where $t_{x}^{\prime}+t_{y}^{\prime}%
i=\rho^{-3}(t_{x}+t_{y}i)=-t_{y}+t_{x}i$ and $d_{x}^{\prime}+d_{y}^{\prime
}i=\rho^{-3}(d_{x}+d_{y}i)=-d_{y}+d_{x}i$. Moreover, the actions in the
\textquotedblleft then\textquotedblright\ and \textquotedblleft else
\textquotedblright cases (as well as the tree indexes) are to be mapped by
$\rho^{3}$. Therefore, if the transient node is in $B_{2}$ and the
destination node is in $Q_{4}$, the action to be performed is given by the
conditional statement \textquotedblleft If $-t_{y}=-k-1-d_{y}$ then
$-1^{\left[  2\right]  }$ else $i^{\left[  2\right]  }$\textquotedblright.
\end{example}

\begin{lemma}
The trees $T_{k}^{\left(  j\right)  },j=1,2,3,4$ are pairwise independent.
\end{lemma}

\begin{proof}
The independence follows from the fact that the action of a transient node in
each case in Table \ref{routingDirection} (and the corresponding rotations) is
unique.
\end{proof}

The above routing algorithm is simplified in Algorithms \ref{StartRouting} and
\ref{Routing} as follows. Let $S = s_{x} + s_{y}i$ be the source node, $T =
t_{x} + t_{y}i$ be the transient node, $D = d_{x} + d_{y}i$ be the destination
node, and $T^{(j)}_k$, $1 \leq j \leq4$, be the tree used for routing.
Furthermore, let $\rho^{j}$ be the previously defined rotation where $j$ denotes
the number of rotations $mod \ 4$. We define the function degree$_{j}$($C$)
that returns the degree of node $C$ in tree $T^{(j)}_k$. The partitions in Figure
\ref{GaussianPartitions} are considered in the following algorithms.

To send a message from $S$ to $D$ using tree $T^{(j)}_k$, the node $S$ calls
StartRouting($S$, $D$, $j$) as described in Algorithm \ref{StartRouting}.
Since the network is symmetric, initially, the algorithm maps the source node
$S$ to node 0 and all other nodes are assumed mapped accordingly. Then, it starts
the routing process by sending the message through the edge that is connected
to the neighbor node corresponding to the tree $T^{(j)}_k$, $1 \leq j \leq4$. For
example, calling StartRouting($S$, $D$, 2) will send a message from node $S$
to node $S+i$ through the $+i$ edge since the $+i$ edge of node $S$ is the
initial direction of the tree $T^{(2)}_k$.

\begin{algorithm}
\caption{StartRouting($S$, $D$, $j$)}
\begin{algorithmic}[1]
\STATE Map the node $S$ to node 0 and assume all other nodes are mapped accordingly including node $D$
\IF{$j = 1$}
\STATE Send through $+1$ packet ($S$, $D$, $j$)
\ELSIF{$j = 2$}
\STATE Send through $+i$ packet ($S$, $D$, $j$)
\ELSIF{$j = 3$}
\STATE Send through $-1$ packet ($S$, $D$, $j$)
\ELSE
\STATE Send through $-i$ packet ($S$, $D$, $j$)
\ENDIF
\end{algorithmic}
\label{StartRouting}
\end{algorithm}

After that, each receiving node performs the steps described in Algorithm
\ref{Routing} as follows. In lines 1-2, the current node $C$ computes its
address relatively to node $S$ after being mapped. In lines 3-5, every
transient node which receives the packet ($S$, $D$, $j$) sets the variables
\textit{dir} to be the direction of the receiving edge, $j$ to hold the tree
number that is being used for the routing, and $r$ to be the number of
rotations required to set the direction of the forwarding edge. Then, in lines
13-14, the transient node checks weather it is the destination node to consume
the packet. Otherwise, the rest of the algorithm, the transient node forwards
the message through a certain direction (edge) based on its degree and the
location of the destination node. Note that, Algorithms \ref{StartRouting} and
\ref{Routing} can be used for any $S$ being a root by mapping $S$ and all
other nodes accordingly. Since all other nodes are also mapped then the
transient node's addresses are computed relatively to address of node $S$.

\begin{algorithm}
\caption{Routing: Transient node process based the received packet ($S$, $D$, $j$)}
\begin{algorithmic}[1]
\STATE Let $C$ be the current working node of form $x+yi$
\STATE Compute the current node's ($C$) relative address to $S$
\STATE Let $dir$ be the receiving direction
\STATE Let $j$ denotes the tree $T_j$ to be used for the routing
\STATE Let $r$ $\gets$ $4-j+1 \ mod \ 4$ be the number of rotations
\STATE CND1 $\gets$ $\rho^r(D)$ $\in$ $V_k - \{R_1 \cup Q_1 \cup B_3 \cup P_3 \cup S_3\}$ \\
\hspace{\algorithmicindent}\hspace{\algorithmicindent}\hspace{\algorithmicindent}\hspace{\algorithmicindent} \AND \\
\hspace{\algorithmicindent}\hspace{\algorithmicindent}\hspace{\algorithmicindent}\hspace{\algorithmicindent}$F_x(\rho^r(C)) \ mod \ k \neq (F_x(\rho^r(D)) + k) \ mod \ k$
\STATE CND2 $\gets$ $\rho^r(D)$ $\in$ $R_2 \cup Q_2 \cup B_3 \cup P_3 \cup S_3$ \AND \\
\hspace{\algorithmicindent}\hspace{\algorithmicindent}\hspace{\algorithmicindent}\hspace{\algorithmicindent}$F_x(\rho^r(C)) \neq F_x(\rho^r(D)) + k + 1$
\STATE CND3 $\gets$ $F_x(\rho^r(C))+1 = F_x(\rho^r(D))$
\STATE CND4 $\gets$ $\rho^r(D) \in R_1 \cup Q_1 \cup R_3 \cup Q_3 \cup B_4 \cup P_4 \cup S_4$
\IF{$C$ = $D$}
\STATE Consume packet
\ELSE
\IF{degree$_j$($C$) = 4}
\IF{CND1 \OR CND2 \OR CND3}
\STATE Send through $\rho^2(dir)$ packet ($S$, $D$, $j$)
\ELSIF{CND4}
\STATE Send through $\rho^3(dir)$ packet ($S$, $D$, $j$)
\ELSE
\STATE Send through $\rho(dir)$ packet ($S$, $D$, $j$)
\ENDIF
\ENDIF
\IF{degree$_j$($C$) = 3}
\IF{$C$ + $\rho^3(dir)$ = $D$}
\STATE Send through $\rho^3(dir)$ packet ($S$, $D$, $j$)
\ELSE
\STATE Send through $\rho^2(dir)$ packet ($S$, $D$, $j$)
\ENDIF
\ENDIF
\IF{degree$_j$($C$) = 2}
\STATE Send through $\rho^2(dir)$ packet ($S$, $D$, $j$)
\ENDIF
\ENDIF
\end{algorithmic}
\label{Routing}
\end{algorithm}

Since Algorithm \ref{StartRouting} is used only once by the source node, its
total communication overhead is 1. Further, the local computation in the
algorithm is based on 5 elementary operations as follows: 1 operation for
mapping the node $S$, at most 3 operations for checking the conditions, and 1
operation for sending the packet when a condition is satisfied. Thus, the
total computation work needed to forward a message is 5 elementary operations
at most.

Algorithm \ref{Routing} is invoked by the rest of the nodes (transient and
destination nodes) in the routing process. The communication overhead for a
single node is 1 since it only forwards 1 packet when a certain condition is
satisfied. Note that, for a path of length $n$, the source node is not counted
since it only performs Algorithm \ref{StartRouting}. Consequently, the total
communication overhead is $n-1$. The local computation of Algorithm
\ref{Routing} for a single node is based on at most 12 operations as follows.
5 of them are assignment operations from line 3 to line 10. 7 operations are
related to the conditions that check the degree of the current node and to
decide in which direction to forward the packet. 1 operation is to send a
packet when a certain condition is satisfied. Thus, since the source node is
not counted, the local computation for a path of length $n$ is $12(n-1)$

\section{Independent Spanning Trees Parallel Construction\label{SectionKConstruction}}
In this section, we present parallel algorithms to construct the four independent spanning trees.
Based on the partition presented in Section~\ref{SectionRouting}, Table \ref{parentChild} shows the parent and children nodes of each node in the first spanning tree. For example, as shown in Figure \ref{T1}, node $-i \in S_4$ has parent node $-2i$ through the edge $-i$ and has no child. Furthermore, node $-1+i \in R_2$ has parent node $-1+2i$ through $+i$ edge and has child nodes $i$ and $-1$ through $+1$ and $-i$ edges, respectively.

\begin{table}[]
\centering
\caption{A node parent and children edge directions as per the partition in
Figure~\ref{GaussianPartitions} }
\label{parentChild}
\begin{tabular}{|l|c|c|c|c|}      \hline
Node in & Parent & Child       \\ \hline
$B_1$   & $-1$   & $+1, \pm i$ \\ \hline
$R_1$   & $-i$   & $+i$        \\ \hline
$Q_1$   & $-i$   & $+i$        \\ \hline
$P_1$   & $-1$   & $+1, \pm i$ \\ \hline
$S_1$   & $-1$   & $+1, \pm i$ \\ \hline

$B_2$   & $-1$   & --          \\ \hline
$R_2$   & $+i$   & $+1, -i$    \\ \hline
$Q_2$   & $+i$   & $-i$        \\ \hline
$P_2$   & $-1$   & --          \\ \hline
$S_2$   & $-1$   & --          \\ \hline

$B_3$   & $+i$   & --          \\ \hline
$R_3$   & $-i$   & --          \\ \hline
$Q_3$   & $-i$   & $+i$        \\ \hline
$P_3$   & $+i$   & --          \\ \hline
$S_3$   & $+i$   & --          \\ \hline

$B_4$   & $-i$   & $+i$        \\ \hline
$R_4$   & $+i$   & $-i$        \\ \hline
$Q_4$   & $+i$   & $-i$        \\ \hline
$P_4$   & $-i$   & $+i$        \\ \hline
$S_4$   & $-i$   & --          \\ \hline
\end{tabular}
\end{table}

In order to get the tables related to the other trees ($T^{(2)}_k$, $T^{(3)}_k$, $T^{(4)}_k$), we define  a $90^o$ counterclockwise rotation mapping $\sigma$ on Table \ref{parentChild} as $\sigma$(Table \ref{parentChild}) = ($\delta$(Node in), $\rho$(Paernt), $\rho$(Child)). The $\delta$ is a cyclic 5-shift on the "Node in" column and $\rho$ is the previously defined rotation. Let the column "Node in" $= (B_1, R_1, Q_1, P_1, S_1, B_2, R_2, \dots , P_4, S_4)$, the cyclic 5-shift $\delta$ on column "Node in" is $\delta$(Node in) $= (B_2, R_2, \dots , P_4, S_4, B_1, R_1, Q_1, P_1, S_1)$. That is, we move the first 5 entries of the first column to the end of the same column. Thus, the parent and child nodes of each node in trees $T^{(2)}_k$, $T^{(3)}_k$, and $T^{(4)}_k$ are $\sigma$(Table \ref{parentChild}), $\sigma^2$(Table \ref{parentChild}), and $\sigma^3$(Table \ref{parentChild}), respectively. The $\sigma^t$ means that the rotation is applied $t$ times on Table \ref{parentChild}.

The following two parallel algorithms constructs the four independent spanning trees. Algorithm \ref{parallel4ISTRoot} triggers the parallel constructions of the trees from root $S$.

\begin{algorithm}
\caption{RootK($S$): Parallel construction of four node independent spanning trees form a source node $S$}
\begin{algorithmic}[1]
%\STATE Map the source node $S$ to node 0 and all other nodes are mapped accordingly.
\STATE addr $\gets$ address of $S$.
\STATE $S$ sends through $+1$ packet (addr)
\STATE $S$ sends through $+i$ packet (addr)
\STATE $S$ sends through $-1$ packet (addr)
\STATE $S$ sends through $-i$ packet (addr)
\end{algorithmic}
\label{parallel4ISTRoot}
\end{algorithm}

In Algorithm \ref{parallel4ISTIntermediate}, we use a static variable as per the C programming language semantics. We assume that each node invokes Algorithm \ref{parallel4ISTIntermediate} as an independent local function. A static variable is declared and initialized only once at the local function first invocation. Its lifetime extends till the global termination of the parallel construction, and it preserves its value between different invocations. Using a static variable enables deciding whether a node has been visited or not as in Algorithm \ref{parallel4ISTIntermediate}, steps 1 to 5. If the node has been already visited, the current node simply ignore the received packet. Otherwise, the current node computes its relative address based on the received packet (addr). Then, it matches its relative address with the entries of Table \ref{parentChild} and $\sigma^t$(Table \ref{parentChild}), for $t = 1, 2, 3$, to determine the edge directions of its parents and children in all spanning trees. After that, it forwards the received packet (addr) to all its neighbor nodes except the one it has already received it from.

\begin{algorithm}
\caption{Intermediate node process based on the received packet (addr)}
\begin{algorithmic}[1]
\STATE Define a static variable $a = 0$.
\IF{$a \ne 0$}
    \STATE Exit as this node already has been visited.
\ENDIF
\STATE $a = a + 1$.
\STATE Compute the relative address of the current node based on the received addr.
\STATE Match the current node relative address with the entries of Table \ref{parentChild} and $\sigma^t$(Table \ref{parentChild}), for $t = 1, 2, 3$, to determine the current node parents and children links in all independent spanning trees.
\STATE Send the addr to all neighbor nodes except the one it was already received from.
\end{algorithmic}
\label{parallel4ISTIntermediate}
\end{algorithm}

We will derive the algorithms' number of steps assuming that
each node can simultaneously send and receive on all its edges.
If the network is fault-free, the algorithms construct the four trees in~$k+1$ steps.
This follows from the facts that the trees construction propagates in all directions
and the network diameter is $k$. Thus, each node is reached in at most $k$ steps.
The nodes that receive the packet in the $k^{th}$ step will forward
it to their neighbors, and this constitutes the one extra step.

If there are one to three node failures, the algorithms construct
at least one fault-free path between each node in the network and the root node.
In this case, however, the trees are not necessarily constructed
as a faulty node could split a spanning tree into two subgraphs.
The construction is bounded by $2k+1$ steps.
The trees construction propagates in all directions, and hence each tree
structure is traced.
By Lemma~\ref{height}, each tree height is
$2k$. Thus, each node will be reached through at least one tree in at most $2k$ 
steps. The nodes that receive the packet in the $2k^{th}$ step will forward
it to their neighbors, and this adds the one extra step.

The total messages generated by the construction algorithms is $6k^2 + 6k + 4$ as
each node generates three messages except the root node generates four.
%Note that this is independent whether the network is fault-free or not.

Algorithm \ref{parallel4ISTRoot} runs in the root node only. Its local computation is limited to 5 operations as follows: one operation to assign the $addr$ variable and 4 operations to send a message to all neighbor nodes.
Except the root node, all network nodes execute Algorithm \ref{parallel4ISTIntermediate} whose local computations is around 15 operations. Thus, the total amount of computation work is
$15(2k^2 + 2k)+5 = 30k^2 + 30k + 5$.

\section{Simulation\label{SectionSimulation}}

\begin{figure}[H]
    \centering
    \includegraphics[scale=0.5]{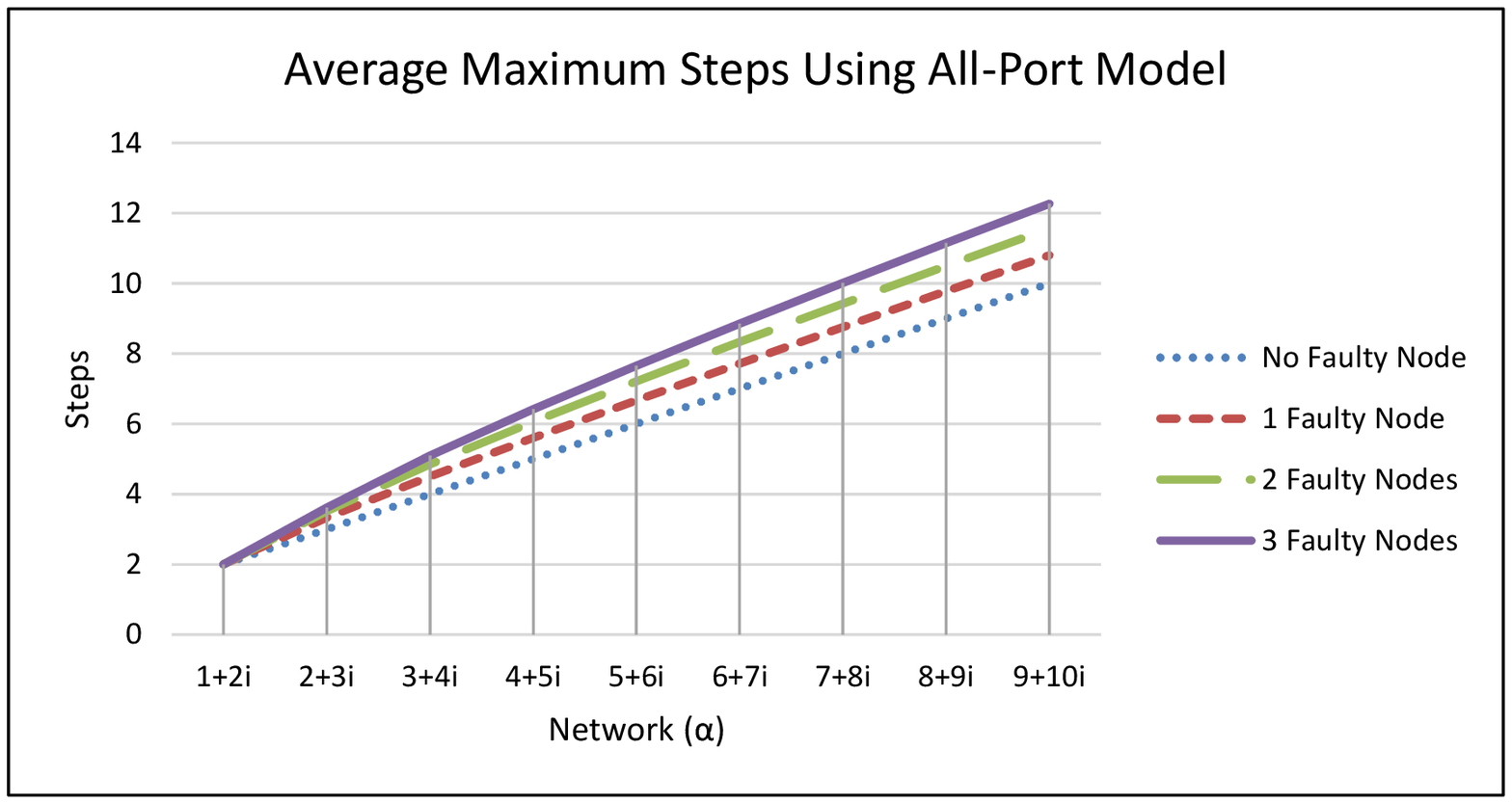}
    \caption{Average maximum steps.}
    \label{allPortsAvgMaxSteps}
\end{figure}

This section discusses the simulation results of our study. We analyzed the construction of independent spanning trees based on the following cases: No faulty node, 1 faulty node, 2 faulty nodes, and 3 faulty nodes in the network. We assumed that each node can simultaneously send and receive to all its neighbor nodes. We have simulated all possible faulty node combinations and measured the average of all maximum steps to construct the trees or paths in different networks as shown in Table \ref{avgMaxStepsAllPort} and Figure \ref{allPortsAvgMaxSteps}. We also measured the maximum of all maximums as displayed in Table \ref{maxStepsAllPorts}.

The simulation results are consistent with the bounds we derived in the previous section.

\begin{table}[H]
\centering
\caption{Average maximum number of steps to construct all trees or paths.}
\label{avgMaxStepsAllPort}
\resizebox{\columnwidth}{!}{
\begin{tabular}{|c|c|c|c|c|c|c|c|c|c|}
\hline
$\alpha$  & 1+2i & 2+3i  & 3+4i  & 4+5i  & 5+6i  & 6+7i  & 7+8i   & 8+9i   & 9+10i  \\ \hline
No Faulty & 2    & 3     & 4     & 5     & 6     & 7     & 8      & 9      & 10     \\ \hline
1 Faulty  & 2    & 3.333 & 4.5   & 5.6   & 6.666 & 7.714 & 8.75   & 9.777  & 10.8   \\ \hline
2 Faulty  & 2    & 3.515 & 4.847 & 6.061 & 7.213 & 8.329 & 9.421  & 10.498 & 11.563 \\ \hline
3 Faulty  & 2    & 3.618 & 5.094 & 6.417 & 7.658 & 8.849 & 10.009 & 11.145 & 12.266 \\ \hline
\end{tabular}
}
\end{table}

\begin{table}[H]
\centering
\caption{Maximum of all maximums number of steps to construct all trees or paths.}
\label{maxStepsAllPorts}
\resizebox{\columnwidth}{!}{
\begin{tabular}{|c|c|c|c|c|c|c|c|c|c|}
\hline
$\alpha$  & 1+2i & 2+3i & 3+4i & 4+5i & 5+6i & 6+7i & 7+8i & 8+9i & 9+10i \\ \hline
No Faulty & 2    & 3    & 4    & 5    & 6    & 7    & 8    & 9    & 10    \\ \hline
1 Faulty  & 2    & 4    & 6    & 8    & 10   & 12   & 14   & 16   & 18    \\ \hline
2 Faulty  & 2    & 4    & 6    & 8    & 10   & 12   & 14   & 16   & 18    \\ \hline
3 Faulty  & 2    & 4    & 6    & 8    & 10   & 12   & 14   & 16   & 18    \\ \hline
\end{tabular}
}
\end{table}

\section{Conclusions\label{SectionConcl}}
In this paper, we presented constructions of four symmetric node independent spanning trees in Gaussian networks, and proved their height is $2k$. Using these trees, we designed routing algorithms that can be used in fault-tolerant and/or secure message communication applications. We also presented fault-tolerant parallel construction algorithms for the presented trees.
These algorithms require $k+1$ steps if the network is fault-free and $2k+1$ steps if one to three faulty nodes exist. The total communication overhead of these algorithms is $6k^2+6k+4$, and the total amount of computation work is $30k^2+30k+5$. We simulated the constructions of the trees in fault-free and faulty networks. The simulation analysis is consistent with our theoretical analysis.

In future research we plan to investigate the constructions of completely independent spanning trees in Gaussian networks and similar regular topology.

\bibliographystyle{IEEEtranS}
\bibliography{4IST}

\end{document}